\documentclass[12pt]{article}
\usepackage{epsf}
\usepackage{amsmath}
\usepackage{graphics}
\usepackage{cite}

\renewcommand{\thefootnote}{\#\arabic{footnote}}
\begin{document}

\newcommand{\gtrsim}{ \mathop{}_{\textstyle \sim}^{\textstyle >} }
\newcommand{\lesssim}{ \mathop{}_{\textstyle \sim}^{\textstyle <} }

\newcommand{\rem}[1]{{\bf #1}}

\renewcommand{\thefootnote}{\fnsymbol{footnote}}
\setcounter{footnote}{0}
\begin{titlepage}

\def\thefootnote{\fnsymbol{footnote}}

\begin{center}
\hfill hep-th/0508082\\
\hfill August 2005\\
\vskip .5in
\bigskip
\bigskip
{\Huge \bf Testable and Untestable Aspects of Dark Energy}
\vskip .45in
\vskip .45in
{\large \bf Paul H. Frampton}

\vskip .45in

{\em
Department of Physics and Astronomy,\\
University of North Carolina, Chapel Hill, NC 27599-3255, USA
}
\end{center}
\vskip .4in
\begin{abstract}
It has been suggested that dark energy will lead to a frequency
cut-off in an experiment involving a Josephson junction.
Here we show that were such a cut-off detected,
it would have dramatic consequences including the possible demise of the string landscape.
\end{abstract}
\end{titlepage}

\renewcommand{\thepage}{\arabic{page}}
\setcounter{page}{1}
\renewcommand{\thefootnote}{\#\arabic{footnote}}

\newpage

.

\vskip 3.0in

{\it Introduction.} ~
The discovery that the expansion of the universe is accelerating
was first made by observations of Type 1A supernovae
\cite{perlmutter,schmidt} and confirmed by independent observations
of the cosmic microwave background (CMB),
especially by the WMAP data\cite{WMAP}.
Assuming general relativity is valid at all length scales, they 
conclude that about 70\% of the energy density of the
universe is in the form of dark energy.

The only successful method of detecting the accelerated expansion
is through such cosmological observations. It has been claimed\cite{beck}
that an experiment involving a Josephson junction
can detect the effect of dark energy by observing a cut-off
in the frequency of zero-point oscillations at about 
1.7 TeraHertz = $1.7 \times 10^{12}$ Hz.
The purpose of the present article is to examine what would be 
consequences for theory if such a cut-off were unexpectedly observed.

\bigskip
\bigskip

{\it Assumptions and Dark Energy.} ~
In an earlier paper\cite{PHF}, it was shown that under three assumptions,
any or all of which may be wrong,
effects of dark energy are in evidence only for length scales in excess
of galactic size, say, 100 kpc. It could possible effect collisions
of two galaxies or larger objects, but not systems of laboratory size
many orders of magnitude smaller.

The assumptions are:

{\bf A.} There exists a stable vacuum with zero cosmological constant. 

{\bf B.} The physical vacuum decays into it by a first-order phase transition.

{\bf C.} There exists a feeble coupling between dark energy and
the electromagnetic field.

Assumptions {\bf A.+B.} are expected to be true in the string landscape
\cite{landscape,landscape2} where the vacuum {\bf A.} is supersymmetric.

\bigskip
\bigskip

{\it Condensed Matter Experiment.} ~
The technical details of the proposed experiment\cite{beck,koch} do not concern us,
although I am assuming that it can measure zero-point fluctuations
as advertized. All I need to
discuss are length scales, and for simplicity of the algebra let
me use a length 1 mm to characterize the condensed matter experiment.
It will be noticed that this length is some 25 orders of magnitude
below the astronomical scale cited in \cite{PHF}.

Let us, however, assume that the experiment finds
a cut-off. What will it teach us?

First, we review the nucleation argument about a first-order phase
transition under the assumptions {\bf A. + B.}. 

The first-order decay is the Lorentz invariant
process of a hyperspherical bubble
expanding at the speed of light, the same for all
inertial observers.
Let the radius of this hypersphere be R, its energy density be $\epsilon$
and its surface tension be $S_1$. Then according to \cite{PHF,PHF76}
the relevant instanton action is
\begin{equation}
A = -\frac{1}{2} \pi^2 R^4 \epsilon + 2 \pi^2 R^3 S_1
\label{action}
\end{equation}
where $\epsilon$ and $S_1$ are the volume and
surface energy densities, respectively.
The stationary value of this action is
\begin{equation}
A_m = \frac{27}{2} \pi^2 S_1^4 /\epsilon^3
\label{stationaryA}
\end{equation}
corresponding to the critical radius
\begin{equation}
R_m = 3S_1 / \epsilon
\label{Rcritical}
\end{equation}
We shall assume
that the wall thickness is negligible compared to
the bubble radius.
The number of vacuum nucleations in the
past lightcone is estimated
as
\begin{equation}
N = (V_u \Delta^4) exp ( - A_m)
\label{nucleations}
\end{equation}
where $V_u$ is the 4-volume of the past and
$\Delta$ is the mass scale relevant to the
problem.  We need
to estimate the three mass-dimension parameters
$\epsilon^{1/4}, S_1^{1/3}$ and $\Delta$ therein
and so we discuss these three scales in turn.

The easiest of the three to select is $\epsilon$. If we imagine a
tunneling through a barrier between a false vacuum
with energy density $\epsilon$ to a true vacuum at energy density zero
then the energy density inside the bubble will
be $\epsilon = \Lambda = (1 mm)^{-4}$. No other
choice is reasonable.
As for the typical mass scale $\Delta$ in the prefactor
of Eq. (\ref{nucleations}), let us put (the reader
can check that the
conclusions do not depend sensitively on this choice)
$\Delta = \epsilon^{1/4} = (1 mm)^{-1}$ whereupon the
prefactor in Eq.(\ref{nucleations}) is $\sim 10^{116}$.
The third and final scale to discuss
is the surface tension, $S_1$, which is here fixed by the assumed
1 mm scale of the experiment. Ignoring order-one numerical factors,
Eq.(\ref{Rcritical}) gives $S_1 = (\epsilon)^{3/4} = (1 mm)^{-3}$.

With these inputs, one finds from Eq.(\ref{stationaryA})
that $A_m = 27 \pi^2 /2$ so that $exp(-A_m) = 10^{-58}$.
Combining this in Eq.(\ref{nucleations}), one finds
$N \sim 10^{58}$ for the number of nucleations meaning that
the dark energy would have decayed gigayears ago.

\bigskip
\bigskip

{\it Discussion.} ~
In order to sufficiently stabilize the dark energy for consistency
with observations, one needs $N << 1$ in Eq.(\ref{nucleations})
and to accomplish this requires the introduction of a dimensionless coupling
between dark energy and the electromagnetic field in assumption {\bf C.}
which is much smaller that $10^{-58}$.

If the coupling were this weak, the experiment has no practicle possibility
of detecting dark energy. The conclusion is that, were the experiment
to make a successful detection, assumptions 
{\bf A. + B.} are untenable. These hold for the
string landscape which would therefore be ruled out. If the
landscape in an inescapable consequence of 
string theory, it would also be excluded.

\bigskip
\bigskip

{\it Falsification of String Landscape.} ~
In the most recent annual conference on string theory
there was a panel discussion viewable at \cite{strings2005}. One
question from the audience was whether there exists any experiment which
could falsify the theory. No clear answer was provided. 
One answer I submit is the experiment proposed in \cite{beck}.

We are ``in the dark" with respect to both dark energy and 
string theory. One attempt to relate the two was  \cite{BFM}.  

Only experiment is a reliable guide so I think it is 
certainly worth checking whether a
cut-off at about 1.7 THz does exist in the Josephson junction experiment.
If such a cut-off were discovered, the consequences would be far reaching
including the possible demise of the string landscape.

\bigskip
\bigskip
\bigskip
\bigskip
\bigskip 
\bigskip
\bigskip
\bigskip

\noindent {\it Acknowledgements}

\bigskip

This work
is supported in part by the
US Department of Energy under
Grant No. DE-FG02-97ER-41036.

\end{document}